\documentclass[structabstract]{aa} 

\usepackage{graphicx}
\usepackage{longtable}
\usepackage{natbib}
\bibpunct{(}{)}{;}{a}{}{,}
\usepackage{txfonts}
\begin{document}

\title{Spectral classification of the mass donors in the high-mass X-ray binaries EXO 1722-363 and OAO 1657-415. \thanks{Based on observations carried out at the European Southern Observatory under programme ID 081.D-0073(A)}}

\author{A.B. Mason \inst{1} 
\and J.S. Clark  \inst{1}
\and A.J. Norton   \inst{1}
\and I. Negueruela \inst{2}
\and P. Roche \inst{1,3,4}}

\institute{Department of Physics \& Astronomy, The Open University, Milton Keynes MK7 6AA, UK \and
 Departamento de F\'{\i}sica, Ingenier\'{\i}a de Sistemas y
  Teor\'{\i}a de la Se\~{n}al, Universidad de Alicante, Apdo. 99,
  E03080 Alicante, Spain \and
  School of Physics \& Astronomy, Cardiff University, The Parade, Cardiff, CF24 3AA, UK. \and
  Division of Earth, Space \& Environment, University of Glamorgan, Pontypridd, CF37 1DL, UK.}

\date{Received 13 May 2009 / Accepted 09 July 2009 }

\abstract{}
{We report near-infrared (NIR) observations of the mass donors of the eclipsing high-mass X-ray binary (HMXB) systems \object{EXO 1722-363} and \object{OAO 1657-415} in order to derive their accurate spectral classifications.}{ESO/VLT observations of the targets with the NIR spectrometer ISAAC were compared with several published NIR spectral atlases of O and B supergiants, an identification of each object's spectral characteristics was made, enabling the refinement of spectral classification of the mass donors.}{We determined that \object{EXO 1722-363} was of spectral type B0 - B1Ia, positioned at a distance $8.0_{-2.0}^{+2.5}$ kpc with a progenitor mass in the range 30M$_{\odot}$ - 40M$_{\odot}$. Luminosity calculations imply that L$_{\rm X}$ $\sim$ 10$^{35}$ - 10$^{37}$ erg s$^{-1}$ for this distance range. We conclude that \object{EXO 1722-363} shares many of the properties associated with other X-ray binary B-type supergiant donors.
\\ 
We found that \object{OAO 1657-415} correlates closely with the spectra of a class of transitional objects, the Ofpe/WNL stars, an intermediate evolutionary stage between massive O type stars leaving the main sequence and evolving into Wolf-Rayets. Due to the wide range in luminosity displayed by Ofpe/WNL stars, (log(L/L$_{\odot}$) $\sim$ 5.3 - 6.2) distance determinations are problematic. For \object{OAO 1657-415} we report a distance of 4.4 $\le$ d $\le$ 12 kpc, implying an X-ray luminosity of 1.5 $\times$ 10$^{36}$ $\le$ L$_{\rm X}$ $\le$ 10$^{37}$ erg s$^{-1}$. We have used our new classification of \object{OAO 1657-415} to explain the physical processes responsible for its unique position within the Corbet diagram. Ofpe/WNL stars demonstrate a high rate of mass-loss through a dense stellar wind combined with a low terminal velocity. This combination of wind properties leads to a high accretion rate and transfer of angular momentum to the neutron star in this system. We believe this in turn leads to a smaller instantaneous equilibrium spin period with respect to normal OB supergiants.}   
{}


\authorrunning{A.B. Mason et al}
\titlerunning{Classification of the HMXBs EXO 1722-363 and OAO 1657-415} 

\maketitle

\section{Introduction}

\paragraph{}

High-mass X-ray binaries (HMXBs) are comprised of massive ($\ge$ 10M$_{\odot}$) donor stars and an accreting compact object (a neutron star or black hole). They are typically divided into two main classes, Be/X-ray binaries (Be/X) and Supergiant X-ray binaries (SGXRBs). Be/X binaries form the majority ($\sim$ 60$\%$) of HMXB systems \citep{liu06}, consisting of a neutron star which accretes matter from the circumstellar equatorial disc of a Be star. These systems have wide orbits and moderate eccentricities, exhibiting two main types of transient X-ray outbursts. Type I occur at the periastron passage of the neutron star with L$_X$ $\sim$ 10$^{36}$ - 10$^{37}$ erg s$^{-1}$. Type II outbursts, which are not correlated with the orbital period, display luminosities of L$_X$ $\ge$ 10$^{37}$ erg s$^{-1}$ \citep{Okazaki01}. SGXRBs have counterparts which are early type supergiants and accrete from either Roche-lobe overflow or a radially outflowing stellar wind. They are persistent sources of X-ray emission, with stellar wind fed systems having a lower flux than Be/X systems (L$_X$ $\sim$ 10$^{35}$ - $10^{36}$ erg s$^{-1}$). For stars that fill their Roche lobe a much higher X-ray luminosity can be achieved of $\sim$   10$^{38}$ erg s$^{-1}$  \citep{liu06}. \\
The Corbet diagram \citep{corbet86} describes the relationship between orbital period $P$$_{\rm orb}$ and pulse period $P$$_{\rm s}$ for an accretion powered pulsar in a HMXB system. The correlation $P$$_{\rm s}$ 
$\propto$ $P_{\rm orb}^{4/7}$ for wind fed systems was found, contrasting with that for Be/X systems of $P$$_{\rm s}$ $\propto$ $P$$_{\rm orb}^2$. Each seperate class lies in a distinct location on the Corbet diagram allowing a differentiation to be made between Be/X and both underfilled and filled Roche-lobe SGXRBs systems. 
 
 Eclipsing X-ray pulsar systems provide a means to accurately determine the mass of the neutron star. Such systems are of significant importance, as they are the only binary accreting system in which the neutron star mass may be measured, providing insights and constraints on the neutron star equation-of-state.
Unfortunately, only ten eclipsing HMXB systems have been identified within the Galaxy, meaning that the characterisation of further examples is a priority. In this paper we undertake the first step in this process for two systems in which the neutron star masses have yet to be measured; the classification of the mass donors in  \object{EXO 1722-363} and \object{OAO 1657-415}.

\subsection{EXO 1722-363}
\object{EXO 1722-363} (\object{IGR J17252-3616}) was discovered in 1984 by $EXOSAT$ Galactic plane observations \citep{warwick88}. Further observations carried out in 1987 by the $Ginga$ satellite showed the presence of pulsations with a  413.9 $\pm$  0.2s period. The X-ray source was found to be highly variable with the 6 - 21 keV flux decreasing from 2 mcrab to 0.2 - 0.3 mcrab
over an 8hr period, with the flux persistent in the 20-60 keV band, but undetectable above 60 keV. At maximum flux the luminosity was found to be 5 $\times$ 10$^{36}$ erg s$^{-1}$ (assuming a distance of 10 kpc, \citet{tawara89}).  The orbital period was refined to $9.741 \pm 0.004$ days and the system eclipse duration was determined as 1.7 $\pm$ 0.1 days \citep{corbet05}.
 Assuming that the donor underfills its Roche lobe and using the eclipse time measurements and the orbital solution from \citet{corbet05}, implies a donor radius between 21 and 37 R$_\odot$ with a mass less than 22 M$_\odot$, and a calculated distance of between 5.3 and 8.7 kpc \citep{thompson07}. From this radius and mass range the donor star was proposed to be a supergiant of spectral type B0 I - B5 I. Observations by $XMM-Newton$ allowed a precise determination of the source positon (to within 4$^{\prime \prime}$) and an IR counterpart 2MASS J17251139-3616575, (with magnitudes J = 14.2, H = 11.8 and K$_{\rm s}$ = 10.7) was proposed independently by both \citet{zurita06} and \citet{negueruela07}.   


\subsection{OAO 1657-415}
\object{OAO 1657-415} was first detected by the $Copernicus$ satellite \citep{polidan78} and later observations with $HEAO-1$ determined a pulse period of 38.22s \citep{white79}. \object{OAO 1657-415} was subsequently found to exist          in an eclipsing  $\sim$10.4~day binary orbit \citep{chak93}. Limitations imposed by  the uncertain spectral classification of the mass donor complicate estimates of the  distance and hence bolometric X-ray luminosity, although \citet{audley} report a distance of 7.1$\pm$1.3~kpc based on observations of the dust-scattered X-ray halo. Nevertheless, the X-ray properties  \object{OAO1657-415} mark it as highly atypical. 
Its location in the Corbet diagram (Fig. 3) separates it from both SG and Be XRBs, implying that it is transitioning from direct wind fed to disc  mediated accretion. 
This hypothesis is supported by a long term secular  spin up of the pulsar on a timescale of $\sim$125~yr (with the 
superposition of additional  brief spin-down and  up episodes, \citet{barnstedt}); the rapidity of 
the process arguing for a short lived phase of stellar/binary evolution. 
  
 The longstanding question of whether \object{OAO 1657-415} was a high-mass or low-mass system was resolved by the {\em Compton Gamma-Ray Observatory} (CGRO) BATSE instrument, deducing from the X-ray pulsar's orbital parameters that the mass donor has a mass of 14-18M$_{\odot}$ with radius 25-32R$_{\odot}$ corresponding to a B0-6 supergiant \citep{chak93}. 
Subsequent examination by the {\em Chandra X-Ray Observatory} determined the precise position of \object{OAO 1657-415} within an error radius of 0.5$^{\prime \prime}$. Optical imaging of the field did not detect any stars at the $Chandra$ identified position upto a limit of V $>$ 23. This implied that the companion experienced significant reddening, requiring near infrared observations to reveal the infrared counterpart star coincident with the $Chandra$ position. The IR counterpart 2MASS J17004888-4139214, (with magnitudes J = 14.1, H = 11.7 and K$_{\rm s}$ = 10.4) was found to have A$_{V}$ = 20.4 $\pm$ 1.3 and a distance of 6.4 $\pm$ 1.5 kpc \citep{chak02}. 

 Near-infrared photometry observations were consistent with the previously reported classification of a B0-B6 supergiant. However, infrared photometry cannot provide an entirely reliable classification, as spectral type and reddening become degenerate on infrared colour-colour diagrams \citep{chak02}. It is possible to perform reliable spectral classification using infrared spectroscopy, with the recent publication of high S/N and resolution spectral atlases of early O and B stars \citep{hanson05} and the advent of IR spectroscopy on large telescopes; interest in this technique has increased as more previously inaccessible objects are classified.    

\section{Observations and data reduction}

Given the relative faintness of both stars (K$\sim$ 10.7 \& 10.4 for \object{EXO 1722-363} and \object{OAO 1657-415}
respectively) we utilised VLT/ISAAC to obtain high S/N and resolution (R$\sim$3000) spectra.
The observations were made on 2008 May 17th in the SW MRes mode with a 0.8 $^{\prime \prime}$ slit. 
To achieve spectral coverage from 2.0-2.22 $\mu$m  two exposures were  obtained, centred at 2.06 $\mu$m 
and 2.15 $\mu$m. Total integration times were  2240s for both \object{EXO 1722-363} and \object{OAO 1657-415}, 
with the resulting data having a count rate below 10,000 ADU; therefore  no correction for non-linearity was 
necessary. Spectra were pipeline reduced and were  wavelength calibrated with OH lines. Finally, 
telluric correction was made utilising two B5V stars, Hip085008 and Hip087805. Unfortunately, due to thin cirrus cloud and highly variable sky conditions, residuals are still present in the vicinity of the He {\sc i} 2.058  $\mu$m line in both spectra
 and the Br$\gamma$ line for \object{OAO 1657-415}. Consequently, while we are confident in the identification 
of these features - noting they are present in the uncorrected target spectra and hence not spuriously 
introduced by division by the telluric standards - we caution against over interpretation of the detailed 
line profiles. We present the spectra of both stars in Figs. 1 \& 2


\section{\object{Spectral Classification of EXO 1722-363}}
In Fig. 1. the spectrum of \object{EXO 1772-363} is compared to template O and B supergiants from \citet{hanson05}. All the absorption lines in this spectrum are narrow, indicative of the object being a supergiant. \object{EXO 1722-363} shows the singlet He\,{\sc i} 2.058 $\mu$m line in emission which is highly sensitive to wind and temperature properties, appearing in absorption in mid to late O stars whilst frequently appearing in emission in early B supergiants \citep{hanson05}. The lack of any lines due to the C {\sc iv} triplet (2.069, 2.078 and 2.083 $\mu$m)  and observed He {\sc i} 2.058 $\mu$m emission implies that \object{EXO 1722-362} is not an O type supergiant. The He {\sc i} 2.058 $\mu$m line in emission contrasts with that of He {\sc i} 2.112 $\mu$m in absorption, this is typical of B0-B2 supergiants \citep{rahoui08}. The N {\sc iii} 2.115 $\mu$m line in emission is a feature common to B0-B1 supergiants. 
 Further evidence pointing to an early B-type classification is the presence of the He {\sc i} 2.184 $\mu$m line, seen here in absorption, combined with the lack of an observed He {\sc ii} 2.188 $\mu$m absorption line, typically absent in spectral types later than O9 \citep{hanson05}.
The lack of strong Br$\gamma$ 2.1655 $\mu$m emission and the absence of Fe\,{\sc ii} 2.089 $\mu$m and Mg\,{\sc ii} 2.138 and 2.144 $\mu$m emission indicates that \object{EXO 1722-363} lacks a strong stellar wind. From a qualitative comparison of spectra from \citet{hanson05}, we identify \object{EXO 1722-363} as being of spectral type B0-B1 Ia. Although we note that we cannot precisely define the luminosity sub-class for two main reasons, firstly the spectral atlas of \citet{hanson05} exhibits a paucity of spectra covering luminosity sub-classes lower than BIa, additionally it is difficult to distinguish between luminosity sub-classes in the K band as the spectral features observed are not highly dependant on luminosity.

 From recent studies of the physical and wind properties of early B supergiants \citep{crowther06}, we find this spectral range has parameters : 22 kK $\le$  T$_{\rm eff}$ ~ $\le$ 28 kK, 22$_{\odot}$ $\le$ R$_{*}$ ~ $\le$ 36$_{\odot}$, 5.35 $\le$ log(L/L$_\odot$) $\le$ 5.65. By comparison with evolutionary rotational massive star models \citep{meynet} we find an initial progenitor mass for \object{EXO 1722-363} in the range 30M$_{\odot}$ - 40M$_{\odot}$ 
The mass we have calculated is based on the object's original progenitor mass. However, with B supergiants experiencing mass-loss at rates M$_{\odot}^{-6}$ yr$^{-1}$ and above \citep{crowther06}, \object{EXO 1722-363} current mass will be significantly reduced from the value we have calculated here. 

 Following the method for determining spectroscopic distance as detailed in \citet{bibby08} and using the parameters shown in Table 1, we determined a distance for \object{EXO 1722-363} of $8.0_{-2.0}^{+2.5}$ kpc which is comparable within errors to the distance deduced in \citet{thompson07}. The quoted uncertainity in the distance estimate principly stems from the uncertainty in absolute magnitude M$_{K_{\rm s}}$  (where we have adopted the mean values in the range of the B0 - B1 I M$_{K_{\rm s}}$ magnitudes from \citet{bibby08}) and interstellar extinction A$_{K_{\rm s}}$.     
Further work will provide a more accurate and detailed mass estimate from on-going radial velocity studies based upon the Br{\sc $\gamma$} line, this will enable additional refinements to be made to distance and luminosity parameters. 

 Comparing our calculated distance with model fluxes derived from spectral fits to \object{EXO 1722-363} \citep{corbet05}, we found that \object{EXO 1722-363} has an intrinsic X-ray flux variability (in the range 2-60 keV) such that F$_{\rm min}$ = 0.78 $\times$ 10$^{-10}$ erg cm$^{-2}$ s$^{-1}$ and F$_{\rm max}$ = 12.2 $\times$ 10$^{-10}$ erg cm$^{-2}$ s$^{-1}$. Using these values combined with the distance range defined above, we derive X-ray luminosities for \object{EXO 1722-363} such that L$_{\rm {X_{min}}}$ = 3.4 $\times$ 10$^{35}$ erg s$^{-1}$ and L$_{\rm {X_{max}}}$ = 1.6 $\times$ 10$^{37}$ erg s$^{-1}$. 
We find this luminosity range entirely consistent with \object{EXO 1722-363} being the donor within an SGXRB system.

\begin{table*}
\caption{\object{EXO 1722-363} H and K magnitudes, with interstellar extinctions A$_{K_{\rm s}}$, adopted absolute magnitudes 
$K_{\rm s}$ for early B supergiants, distance moduli (DM) from \citet{bibby08} with calculated distance.}  
\label{table:2}
\centering
\begin{tabular}{ccccccccc}
\hline
\hline
H & $K_{\rm s}$ & $H - K_{\rm s}$ & $(H - K_{\rm s})_{0}$ & E$_{(H - K_{s})}$ & M$_{K_{\rm s}}$ & A$_{K_{\rm s}}$   & 
DM & Distance \\ 
(mag) & (mag) & (mag) & (mag) & (mag) & (mag) & (mag) & (mag) & kpc \\
\hline
11.81 & 10.67 & 1.14 & $-$0.08 & 1.22 & $-$6.0$\pm$0.5 & 2.2$\pm$0.4 &      14.5$\pm$0.6 & $8.0_{-2.0}^{+2.5}$\\
\hline
\end{tabular}
\end{table*}
   
\begin{figure} [h]
\begin{center}
  \includegraphics[width=8cm]{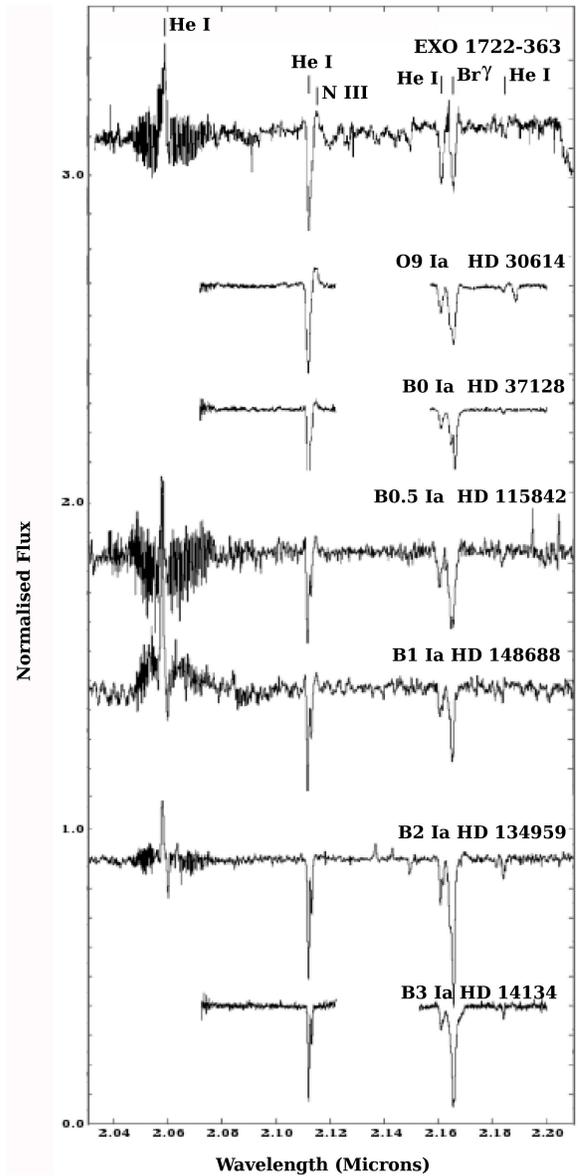}
  \caption{Comparison of \object{EXO 1722-363} and template 09-B3 Ia spectra from \citet{hanson05} in the K band. The fine toothcomb structure at $\lambda$ = 2.06 $\mu$m is due to imperfect telluric correction.}  
\end{center}
\end{figure}

\section{\object{Spectral classification of OAO 1657-415}}

The spectrum of the mass donor in \object{OAO1657-415} is presented in Fig. 2, and is dominated by He\,{\sc i} 
2.058 $\mu$m and Br$\gamma$ emission, the former stronger than the latter. 
He\,{\sc i} 2.112 $\mu$m and the He\,{\sc i}  complex centred around 2.16 $\mu$m are seen in 
aborption, while there is no evidence for high excitation C\,{\sc iv} or low excitation Fe\,{\sc ii} and Mg\,{\sc ii} 
emission features that characterise early-mid O and  mid-late B supergiants respectively. Nevertheless we find a poor 
correspondence with the spectra of B0-6 supergiants \citep{hanson96,hanson05} - as suggested for the 
mass donor by \citet{chak02} on the basis of a combination of photometric and X-ray data.

 However, comparison with the spectra of massive transitional objects presented by  \citet{morris96} is 
more encouraging. Massive transitional stars are a heterogenous grouping of both cool and hot supergiants, 
characterised by extreme mass loss rates which act to remove the H rich mantle of the progenitor  as it evolves 
into a H-depleted Wolf Rayet star. In particular \object{OAO 1657-415} shows pronounced similarities to the hot Ofpe/WNL 
stars. This is dramatically illustrated by comparison to \object{GC IRS16NW}, located within the Galactic  Centre 
cluster \citep{martins07}. Unfortunately, such a classification results in interpretational  difficulties. 

Firstly, such stars  demonstrate a wide range of intrinsic luminosities log(L/L$_{\odot}$) $\sim$ 5.3-6.2; 
(Clark et al., submitted) and hence  progenitor masses ($\sim$25-100+ M$_{\odot}$ respectively). Additionally, such stars also exhibit a known observational degeneracy such that differing combinations of mass loss rate, H/He 
ratio and temperature (and hence luminosity) yield $\sim$identical K band spectral morphologies  
\citep{hillier,martins07}. Consequently we may not {\em a priori} determine a unique distance to 
\object{OAO 1657-415} on the basis of this classification. Adopting the appropriate $(J-K_{\rm s})_o$ colour for such 
stars from \citet{crowthera} and  the reddening law of  \citet{rieke} yields 
A$_{K_{\rm s}}$ = 2.37. Then, assuming a bolometric correction from the non-LTE analysis of \object{GC IRS16NW} by \citet{martins07} we may infer a minimum and maximum distance for \object{OAO 1657-415} by adopting the 
range of intrinsic luminosities given above. In doing so we find inevitably unconstrictive limits of 
4.4~kpc $<$ d $<$ 12~kpc. In turn this results in  1.5 $\times$ 10$^{36}$ ~erg s$^{-1}$ < ~ L$_{\rm X}$ < 10$^{37}$~erg s$^{-1}$, also entirely consistent with  observed luminosities of SG HMXBs. As Ofpe/WNL stars typically demonstrate systematically lower wind velocities and higher mass loss rates than normal OB supergiants (see \S4.2), we would expect to observe a higher than average X-ray luminosity than that typically seen in most other SGXRBs. We believe for this source the X-ray luminosity will approach the upper limit of L$_{X}$ $\sim$ 10$^{37}$~erg s$^{-1}$ we have derived, implying that \object{OAO 1657-415} lies at the upper limits of our quoted distance range.
Alternatively, adopting the distance derived by \citet{audley} leads to log(L/L$_{\odot}$) $\sim$ 5.7. For such a luminosity, comparison to the evolutionary tracks for massive stars from \citet{meynet} 
imply an initial  mass of  $\sim$40~M$_{\odot}$\footnote{Note that in light of the degeneracy described above, 
adoption of a higher stellar temperature would result in a systematic reduction in distance estimates,
 or, for a given distance, an increase in the initial mass of the primary.}.
 We caution that such a mass estimate does not {\em a priori} imply that $\sim$40~M$_{\odot}$ stars yield neutron stars post SNe; for 
example \citet{wellstein} propose a scenario for the B Hypergiant+neutron star HMXB 
\object{GX301-2} in which  conservative case A mass transfer from a 26+25M$_{{\odot}, initial}$ configuration 
leads to the current mass of the donor  ($>$39M$_{\odot}$ \citet{kaper04}).

\begin{figure} [h]
\begin{center}
  \includegraphics[width=8cm]{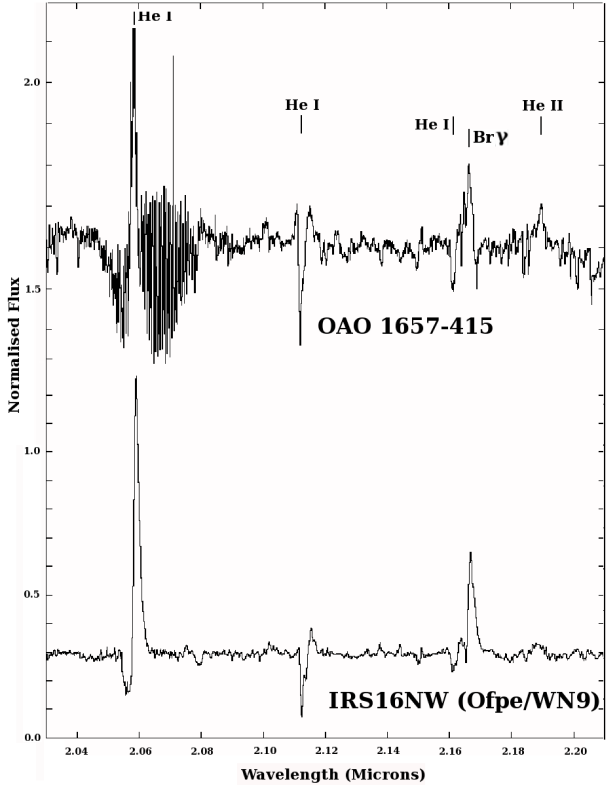}
  \caption{Comparison of OAO 1657-415 and IRS16NW K band spectra from \citet{martins07}.The fine toothcomb structure at $\lambda$ = 2.06 
$\mu$m is due to imperfect telluric correction.}
\end{center}
\end{figure}

\subsection{Formation and  evolution  }

With an Ofpe/WNL primary, \object{OAO 1657-415} adds to the growing number of  HMXB which have mass donors
 in  a more advanced evolutionary state than the canonical OB supergiants of which \object{EXO 1722-363} is an exemplar. 
These include the supergiant B[e] systems \object{CI Cam} \citep{clark99}, \object{IGR J16318-4848} \citep{fill04} \& \object{IGR J16358-4726} \citep{rahoui}, the B hypergiant \object{GX301-2} \citep{kaper04} and the Wolf Rayet star mass donors to \object{Cyg X-3} \citep{kerkwijk} and \object{IC10 X-1}  
\citep{clark04}. Note that the greater number of Galactic HMXBs with canonical OB 
supergiants\footnote{Which, including Supergiant Fast X-ray transients, currently number in excess of 20, with a further 
10 candidates.} likely  reflects a combination of the relatively short  lifetimes of transitional stars and Wolf Rayets with 
respect to OB supergiants and the fact that lower mass stars ($\sim$15-25M$_{\odot}$) - which form viable mass  donors for supergiant HMXBs - are not expected to evolve to such advanced
evolutionary states.

Given the presence of Br$\gamma$ emission in the spectrum of \object{OAO 1657-415}, we caution that it is less evolved than the 
H-free Wolf Rayet mass donors of \object{Cyg X-3} and \object{IC10 X-1}. Hence  it appears  unlikely that it represents the 
post common envelope endpoint of the evolutionary scenario for HMXBs of \citet{heuvel}; 
moreso given that 
such a binary  interaction would be expected to yield a very short orbital period. Additionally, the non zero eccentricity of the 
orbit \citep{chak93} would also argue against a post-SNe common envelope phase. Finally, 
 the {\em Midcourse Souce Experiment} \citep{egan} and the GLIMPSE \& MIPSGAL/{\em Spitzer} legacy surveys
\citep{benjamin,carey} reveal a lack of either point-like or spatially extended 
excess mid-IR emission that would be indicative  of dusty ejecta produced in either a common envelope, Red Supergiant (RSG) or 
Luminous Blue Variable (LBV phase see \S4.2)  \citet{voors,clark03,clark07,fuchs}. Therefore  we consider it
 most likely that the mass donor in \object{OAO 1657-415}  has evolved directly into an Ofpe/WNL phase, implying a massive 
($>$40M$_{\odot}$) progenitor - consistent with a distance of 7.1$\pm$1.3~kpc \citep{audley}, rather than through
 a lower mass (25-40M$_{\odot}$) channel that would require it to be in a {\em post} RSG phase \citep{meynet}.

However, in light of  the absence  of (mid-IR) evidence for significant recent mass loss, this mass estimate 
 appears uncomfortably high given the constraints implied by \citet{chak93} 
on the basis of the X-ray properties: a {\em current} mass for the donor of  14-18M$_{\odot}$ for a 1.4M$_{\odot}$ 
neutron star, with a maximum of $\sim$25M$_{\odot}$ under restrictive circumstances\footnote{The requirement that the mass donor is always within its Roche Lobe, with a low inclination angle and neutron 
star mass.}. We note  that  the mass of the Ofpe/WNL star within the binary \object{Cyg OB2-5} - which is not expected
to have passed through a RSG phase - is also surprisingly low 10.5-15.5M$_{\odot}$ \citep{rauw}. Given these 
observations,  one might speculate that binary mass and  angular momentum transfer to the Ofpe/WNL stars in both systems 
would lead to rapid spin up and hence increased mass loss rates \citep{petrovic}, in turn  leading 
to the current, unexpectedly low stellar masses inferred for both stars. 

\begin{figure}
\begin{center}
  \includegraphics[width=8cm]{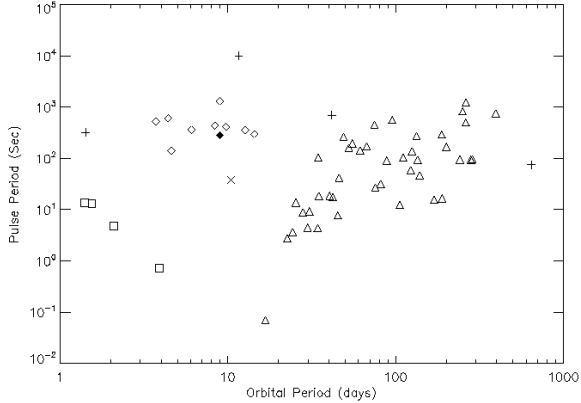}
  \caption{Corbet diagram marking position of OAO 1657-415 and other HMXBs. OAO 1657-415 is marked by an X, EXO 1722-363 by a filled diamond. SGXRB Roche-Lobe Overflow systems  (Squares), Be/X binaries (Triangles), SGXRB Wind-fed systems (Diamonds) and anomalous systems (+)}
\end{center}
\end{figure}

\subsection{X-ray properties }

Finally, we turn to the implications of the Ofpe/WNL classification for the  X-ray properties of \object{OAO 1657-415}. The
 anomalous position of \object{OAO 1657-415} within the Corbet diagram (Figure 3) \citep{corbet86}, is then naturally 
explained in terms of the properties of it's  stellar wind. Compared to normal OB supergiants  
\citep{crowther06}, Ofpe/WNL stars typically demonstrate systematically  lower wind velocities and higher mass loss rates \citep{martins07}.
 This combination of wind properties permits a higher accretion rate and hence  transfer of angular momentum to the 
neutron star, in turn leading to a smaller (instantaneous) equilibrium spin period with respect to normal OB 
supergiants ($P_{\rm spin} 
\propto$ \.{M}$^{-3/7}$$v_{\infty}^{12/7}$ from Eqn. 12 of \citet{waters},  where $P_{\rm spin}$, \.{M} 
and $v_{\infty}$ are the spin period of  the neutron star and the  mass loss rate and terminal velocity of the mass donor
 wind respectively).

Likewise, Ofpe/WNL stars have been proposed to be the hot quisecent state of LBVs - an unstable phase in the 
post-MS lifetime of massive stars which is characterised by dramatic long term  ($\sim$yrs) 
changes in both stellar radius and/or mass loss rate (see \citet{hd} for a review). If 
\object{OAO 1657-415} were an incipient/quiescent LBV, then such increases in either mass loss rate or radius - bringing 
it closer to the Roche Lobe and hence increasing the mass transfer rate - could explain the significant long term ($\sim$months) 
 variability in X-ray luminosity that  it demonstrates \citep{kuulkers}. Indeed, given both  the mass loss rates and 
radial  extent that LBVs have been observed to  reach  (e.g. \.{M} $\geq$ 10$^{-4}$ M$_{\odot}$ yr$^{-1}$
and R$_{*}$ $>$ 100R$_{\odot}$; Clark et al., submitted)\footnote{Significantly larger than the orbital separation of 
\object{OAO 1657-415}, projected semimajor axis {\em a$_x$} $\sin$ $i$ $\sim$ 45R$_{\odot}$, \citet{chak93}} one could anticipate a greatly enhanced mass transfer rate for \object{OAO 1657-415} leading either to an extreme X-ray luminosity or 
the formation of a common envelope and spiral in/merger of both components in a high mass analogue to the RSG common envelope
phase described by \citet{heuvel}. Given such a potential scenario it is of considerable interest 
that the B hypergiant HMXB system \object{GX301-2} shows evidence for  a pronounced circumstellar envelope of the type that is 
indicative of the pronounced mass loss associated with the LBV phase  \citep{moon}.


\section{Conclusions}

Within this paper we have presented the analysis and results of observations performed at ESO/VLT with the ISAAC spectrometer on the eclipsing high-mass X-ray binaries systems EXO 1722-363 and OAO 1657-415. Using NIR spectrometry we have constrained the previous spectral classification of the mass donor in the EXO 1722-363 system from B0-B5 I to B0-BI Ia and determined its distance to be $8.0_{-2.0}^{+2.5}$ kpc.
Examination of the OAO 1657-415 system has allowed us to determine that the donor in this system is more evolved than the typical OB supergiants found in other eclipsing high mass X-ray binaries. We have classified the donor within OAO 1657-415 as type Ofpe/WN9, a transitionary object between OB main sequence stars and hydrogen depleted Wolf-Rayet stars. Due to the large range in luminosity exibited by these types of stars it is difficult to precisely perform distance calculations. Adopting a luminosity range of log(L/L$_\odot$) $\sim$ 5.3 - 6.2 we determined a distance range of 4.4 $\le$ d $\le$ 12 kpc. 

 The anomolous position of OAO 1657-415 on the Corbet diagram is explained by the more evolved (than typical OB supergiant XRBs) mass donor having a stronger, slower stellar wind, enabling the NS to increase its accretion rate and its rate of spin. This result reinforces our belief that the circumstellar environment from which a HMXB accretes from plays a crucial role in determining its X-ray properties. Results from the INTEGRAL $\gamma$-ray telescope \citep{walter06} have lead to the discovery of two new distinct classes of SG XRB within the past decade. Supergiant Fast X-ray Transients (SFXT) \citep{neg06} are believed to stem from periods of high accretion due to wind clumping, and obscured systems such as sgB[e] stars \citep{fill04} have a very high X-ray extinction due to the density of the circumstellar environment. In addition to examining the spin and orbital periods of HMXB systems, it is of vital importance to also consider the properties of the donor's wind. It is increasingly apparent that the X-ray properties of HMXBs are influenced greatly by the circumstellar environment in tandem with the mass-loss properties of the mass donor. We plan to present radial velocity studies of both systems in a future paper, where we shall attempt to determine the masses of the two components in each case.


\begin{acknowledgements}
ABM acknowledges support from an STFC studentship. JSC acknowledges support from an RCUK fellowship. 
This research is partially supported by grants AYA2008-06166-C03-03 and
Consolider-GTC CSD-2006-00070 from the Spanish Ministerio de Ciencia e
Innovaci\'on (MICINN).
Based on observations carried out at the European Southern Observatory, Chile through programme ID 081.D-0073(A). We also thank the anonymous referee for their useful comments.

\end{acknowledgements}

\bibliographystyle{aa}
\bibliography{first_paper}


\end{document}